\documentclass[a4paper]{jpconf}
\usepackage{graphicx}
\usepackage{bm}
\usepackage{cite}
\begin{document}
\title{Magnetic-field dependence of antiferromagnetic structure in CeRh$_{\bm{1-x}}$Co$_{\bm{x}}$In$_{\bm{5}}$}
\author{M Yokoyama$^{1,\dag}$, Y Ikeda$^2$, I Kawasaki$^2$, D Nishikawa$^2$, K Tenya$^3$,\\ K Kuwahara$^4$ and H Amitsuka$^2$}
\address{$^1$Faculty of Science, Ibaraki University, Mito 310-8512, Japan}
\address{$^2$Graduate School of Science, Hokkaido University, Sapporo 060-0810, Japan}
\address{$^3$Faculty of Education, Shinshu University, Nagano 380-8544, Japan}
\address{$^4$Institute of Applied Beam Science, Ibaraki University, Mito 310-8512, Japan}
\ead{$^\dag$makotti@mx.ibaraki.ac.jp}

\begin{abstract}
We investigated effects of magnetic field $H$ on antiferromagnetic (AF) structures in CeRh$_{1-x}$Co$_x$In$_5$ by performing the elastic neutron scattering measurements. By applying  $H$ along the $[1\bar{1}0] $ direction, the incommensurate AF state with the propagation vector of $q_{h1}=(1/2,1/2,0.297)$ observed at $H=0$ is replaced by the commensurate AF state with the $q_{c2} = (1/2, 1/2, 1/4)$ modulation above $2\ {\rm T}$ for $x = 0.23$, while the AF states with the $q_{c1}=(1/2,1/2,1/2)$ and $q_{h2}=(1/2,1/2,0.42)$ modulations seen at $H=0$ change into a single $q_{c1}$-AF state above $\sim 1.6\ {\rm T}$ for $x = 0.7$. These results suggest the different types of AF correlation for Co concentrations of 0.23 and 0.7 in an applied magnetic field $H$. 
\end{abstract}

\section{Introduction}
The heavy-fermion compounds Ce$M$In$_5$ ($M={\rm Co\ and\ Rh}$: HoCoGa$_5$-type tetragonal structure) have been attracting much interest for a rich variety of physical properties originating from the interplay of antiferromagnetism and superconductivity.  CeRhIn$_5$ shows an antiferromagnetic (AF) order below $T_{N}=3.8\ {\rm K}$ \cite{rf:Hegger2000}, whose structure is suggested to be a helical with an incommensurate (IC) propagation vector $q_{h1}=(1/2,1/2,0.297)$ \cite{rf:Bao2000}. It is found that applying pressure $p$ suppresses the IC-AF state, and then induces the superconducting (SC) state above $p=1-1.5\ {\rm GPa}$ \cite{rf:Mito2001,rf:Knebel2006}. In CeCoIn$_5$, on the other hand, the SC state is observed below $T_c=2.3\ {\rm K}$ at ambient pressure\cite{rf:Petrovic2001}. The SC and AF states are revealed to be  effectively controlled by magnetic field $H$: applying $H$ generates the non-Fermi-liquid state above SC critical field $H_{c2}$ \cite{rf:Bianchi2003} and the unconventional SC phase with the IC-AF modulation just below $H_{c2}$ \cite{rf:Young2007,rf:Kenzelmann2008}. These features are considered to occur due to the AF correlations enhanced in the vicinity of the AF quantum critical point (QCP).  

The relationship between the AF and SC phases has first been investigated in CeRh$_{1-x}$Co$_x$In$_5$ by thermodynamic and transport measurements \cite{rf:Zapf2001}. It is found that $T_{N}$ of the IC-AF order is weakly reduced by increasing $x$, and then approaches zero at the QCP: $x_c\sim 0.8$. At the same time, the SC phase develops above $x\sim 0.4$. The proposed $x-T$ phase diagram is similar to the $p-T$ phase diagram for pure CeRhIn$_5$. Recent neutron scattering experiments revealed that a commensurate (C) AF order with a modulation of $q_{c1}=(1/2,1/2,1/2)$ appears in the intermediate $x$ range \cite{rf:Yoko2006,rf:Kawamura2007}. Furthermore, we have observed that the $q_{h1}$ modulation changes into $q_{h2}=(1/2,1/2,0.42)$ at $x\sim 0.7$ \cite{rf:Yoko2008}. These results suggest that the nature of the AF correlation varies by doping Co, and it may significantly affects the evolution of the SC order. To elucidate magnetic instability involved in small and rich Co concentrations, we have investigated the magnetic-field dependence of the AF structure for CeRh$_{1-x}$Co$_x$In$_5$ by performing the elastic neutron scattering experiments.

\section{Experiment Details}
Single crystals of CeRh$_{1-x}$Co$_x$In$_5$ were grown by the In-flux technique, and they were cut into rod shape ($\sim 1.6\times 1.6\times 10\ {\rm mm^3}$) along the $[1\bar{1}0]$ axis. In accordance with the previous investigations \cite{rf:Yoko2008}, the Rh/Co concentrations $x$ in the samples were checked by means of the electron probe microanalysis (EPMA) measurements, and we adopt the $x$ values estimated from the EPMA measurements in this study. The elastic neutron scattering (ENS) experiments for $x=0.23$ and 0.7 were performed on the triple-axis spectrometers ISSP-GPTAS (4G) and ISSP-PONTA (5G) located at the research reactor JRR-3M of JAEA, Tokai. We selected a neutron momentum $k=3.83\ {\rm \AA}^{-1}$ by pyrolytic-graphite (PG) monochromator and analyzer, and used a combination of 40'-40'-40'-80' collimator and two PG filters. The samples were cooled down to 1.6 K in a pumped $^4$He cryostat with a superconducting magnet. The sample rods were set to be perpendicular to the  $(hhl)$ scattering plane. This  scattering geometry enables to minimize an effect of neutron absorption by Rh and In. Magnetic field $\mu_0H$ was applied up to 3 T along the $[1\bar{1}0]$ direction.

\section{Results and discussion} 
\begin{figure}[tbp]
\begin{center}
\vspace{-15pt}
\includegraphics[keepaspectratio,width=0.6\textwidth]{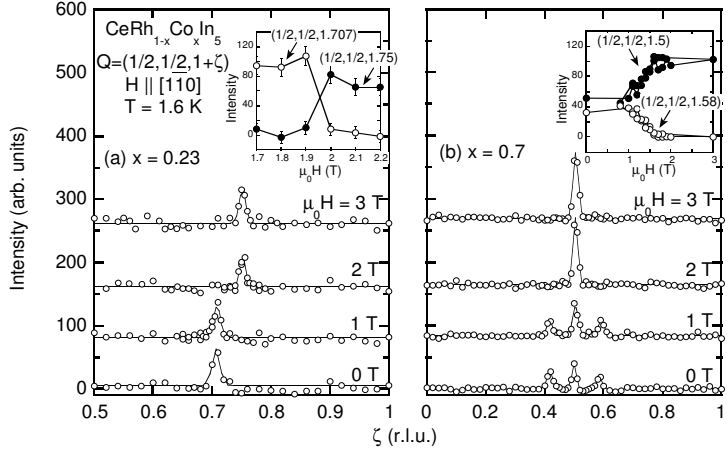}
\vspace{-15pt}
\end{center}
  \caption{
The ENS patterns under magnetic fields at 1.6 K for CeRh$_{1-x}$Co$_x$In$_5$ with (a) $x=0.23$ and (b) 0.7 obtained by scanning at the momentum transfer $Q=(1/2,1/2,1+\zeta)$. The baselines for the data are shifted for clarity. Note that the range of the horizontal axis is enlarged around $\zeta=0.75$ in (a). The insets show the ENS intensities at the AF Bragg-peak positions plotted as a function of $\mu_0H$.
}
\end{figure}
Figure 1 shows the $H$ variations of the ENS patterns at 1.6 K for $x=0.23$ and 0.7, obtained by scanning at momentum transfers $Q=(1/2,1/2,1+\zeta)$. Instrumental backgrounds were carefully subtracted using the data at 5 K $(>T_N)$. At $x=0.23$, a set of Bragg peaks due to the IC-AF order with the modulation of $q_{h1}=(1/2,1/2,0.297(6))$ was observed at zero field \cite{rf:Yoko2008}. Both position and intensity of these Bragg peaks are unchanged under $H$ smaller than $1.9\ {\rm T}$. At 2 T, the IC-AF Bragg peak suddenly moves toward a commensurate position, accompanying a slight reduction of the peak intensity (inset of Fig.\ 1(a)). This indicates an occurrence of a C-AF phase with a propagation vector of $q_{c2}=(1/2,1/2,1/4)$ above 2 T, separated from the $q_{h1}$-AF phase via a first-order transition. Since the same transition is also observed in pure CeRhIn$_5$ \cite{rf:Raymond2007}, the magnetic properties are considered to be unchanged for the small amount of Co doping.  

For $x = 0.7$, the application of $H$ induces commensurate AF state which are different from that observed for $x = 0.23$. The Bragg-peaks for the AF orders with $q_{h2}=(1/2,1/2,0.42)$ and $q_{c1}=(1/2,1/2,1/2)$ appear at $H=0$. They are integrated into the C-AF Bragg peaks with the $q_{c1}$ modulation for field above $\sim 1.3\ {\rm T}$ ($=\mu_0H^*$), {\it i.e.}, a single $q_{c1}$-AF order evolves. Note that the wave vector $q_{c1}$ differs from $q_{c2}=(1/2,1/2,1/4)$ observed for $x=0.23$. In contrast to the sharp transition observed for $x=0.23$, this variation occurs in a wide field range between 1 T and 1.6 T (inset of Fig.\ 1(b)): the Bragg-peak intensity for the $q_{h2}$-AF order linearly decreases in connection with the enhancement of the intensity for the $q_{c1}$-AF order. These Bragg-peak positions do not change in this transition. We suggest that this transition is also of the first order, although in the present experimental accuracy the hysteretic behavior is not detected in the $H$ variations of the Bragg-peak intensities. The total intensity of the these AF Bragg peaks is roughly conserved in the present $H$ range.

\begin{figure}[tbp]
\begin{center}
\vspace{-15pt}
\includegraphics[keepaspectratio,width=0.6\textwidth]{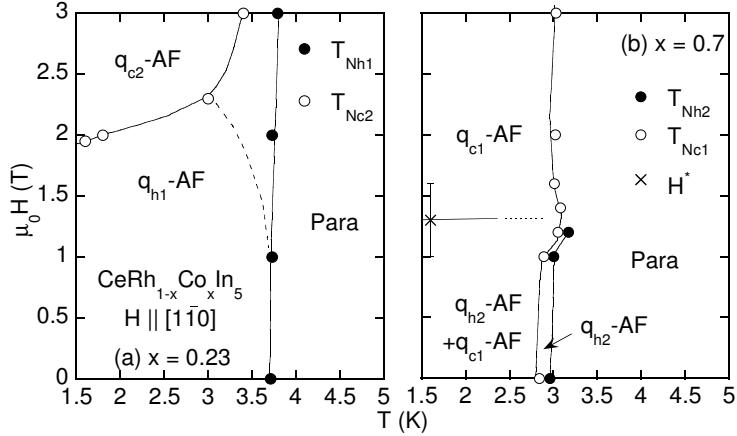}
\vspace{-15pt}
\end{center}
  \caption{
$H-T$ phase diagram for CeRh$_{1-x}$Co$_x$In$_5$ with (a) $x=0.23$ and (b) 0.7, obtained from the temperature and field variations of the AF Bragg-peak intensities. The definitions of $T_{Nh1}$, $T_{Nh2}$, $T_{Nc1}$, $T_{Nc2}$ and $H^*$ are described in the text. In (a), a dashed line indicates a possible phase boundary expected from a comparison with the $H-T$ phase diagram for pure CeRhIn$_5$ \cite{rf:Raymond2007}.
}
\end{figure}
Displayed in Fig.\ 2 are the $H-T$ phase diagrams for $x=0.23$ and 0.7 obtained from the temperature and field scans for the AF Bragg-peak intensities. For $x=0.23$, the onset of the Bragg peaks for the $q_{h1}$-AF order $T_{Nh1}$ slightly increases from $\sim 3.7\ {\rm K}$ at $H=0$ with increasing $H$. At 1.6 K, on the other hand, the C $q_{c2}$-AF phase appears above $\mu_0H=1.9\ {\rm T}$. The boundary between these AF phases estimated from the onset of the $q_{c2}$-AF order $T_{Nc2}(H)$ is extended to high $H$ and $T$ region, and shows a tendency to be parallel with the $T_{Nh1}(H)$ line. This $H-T$ phase diagram is quite similar to that for pure CeRhIn$_5$ \cite{rf:Raymond2007}. In CeRhIn$_5$, an another phase boundary is found at the inside of the $q_{h1}$-AF phase, which can be detected by ENS as a slight variation of the AF Bragg-peak intensity. However, in the present measurement we cannot verify the presence of such boundary for very weak signal from the sample. 

For $x=0.7$, the onsets of the Bragg-peaks for the $q_{h2}$- and $q_{c1}$-AF states ($T_{Nh2}$ and $T_{Nc1}$) are nearly constant at $\sim 3.0\ {\rm K}$ and $\sim 2.8\ {\rm K}$ below $\mu_0H=1\ {\rm T}$, respectively. They show weak enhancements with further increasing $H$. Above $\mu_0H=1.3\ {\rm T}$, $T_{Nc1}$ is slightly reduced, and $T_{Nh2}(H)$ cannot be determined due to the reduction of the $q_{h2}$-AF Bragg peak intensities. Finally, for $\mu_0H > 1.6\ {\rm T}$ the $q_{h2}$-AF state completely vanishes and $T_{Nc1}$ becomes again independent of $H$. The $H$ range of $T_{Nc1}$ being enhanced seems to match with that for the $H$-induced variation of the AF structure at 1.6 K. To obtain precise information on the phase boundaries, we plan to perform the magnetization and specific heat measurements under $H$. 

We have observed that four different AF structures corresponding to the wave vectors of $q_{h1}$, $q_{h2}$, $q_{c1}$, $q_{c2}$ appear  by changing $x$ and $H$ in CeRh$_{1-x}$Co$_x$In$_5$. The evolutions of such various modulations along the $c$ axis are expected to be basically attributed to the two-dimensional characteristics of these compounds. At the same time, it is remarkable that the AF structure with the modulation of $q_{h1}$ are favored in the small $x$ concentrations, while those with $q_{c1}$ and its neighbor are stabilized as $x$ approaches $x_c$. Similar feature is also seen in the $H$ variations of the AF structures: under $H$ the $q_{c2}$-AF phase occurs in the small $x$ range, while $q_{c1}$-AF phase develops at $x \sim x_c$. In most of heavy-fermion systems showing quantum critical behavior, it is considered that enhancement of quantum fluctuation near QCP is ascribed to the suppression of the AF order. Since the QCP is expected to exist at $x_c$ in CeRh$_{1-x}$Co$_x$In$_5$, the change in the AFM order seen in the present experiment close to $x_c$ of 0.7 should be accompanied with fluctuations at the same $q$. Recent de Haas van Alphen experiments for CeRh$_{1-x}$Co$_x$In$_5$ revealed a dramatic change in the topology of the Fermi surface at a much smaller $x$ than $x_c$ \cite{rf:Goh2008}, which is expected to be connected with the variations of the AF structure presently observed.

\section{Summary}
Our ENS experiments under $H$ ($\parallel [1\bar{1}0]$) for CeRh$_{1-x}$Co$_x$In$_5$ revealed that the $H$ variations of the AF structure largely depend on $x$: the application of $H$ changes the IC $q_{h1}$ structure into the C $q_{c2}$ one at $x=0.23$, while it does the $q_{c1}$ and $q_{h2}$ modulations into the single $q_{c1}$ one at $x=0.7$. This suggests that the different types of the AF correlations from those in pure CeRhIn$_5$ are enhanced at around $x_c$, presumably in connection with the variations of the Fermi-surface properties with changing $x$.

\ack
We thank T.J. Sato, M. Matsuura, T. Asami and Y. Kawamura for the technical supports on the ENS experiments.

\end{document}